\documentclass[11pt]{iopart}
\usepackage{iopams,natbib}
\bibpunct{(}{)}{;}{a}{}{,}
 
\pdfoutput=0
\usepackage{graphicx}
\newcommand{\bm}[1]{\mbox{\protect\boldmath$#1$}}
\begin{document}
\title[Influence of frame-dragging on magnetic null points]{Influence of frame-dragging on magnetic null points near rotating black hole}
\author{V Karas, O Kop\'a\v{c}ek and D Kunneriath}
\address{Astronomical Institute, Academy of Sciences, Bo\v{c}n\'{i}~II 1401, CZ-141\,31~Prague, Czech~Republic}
%\ead{vladimir.karas@cuni.cz, kopacek@ig.cas.cz, devaky@ig.cas.cz}

\begin{abstract}
Understanding the mechanisms of particle acceleration from the 
vicinity of black holes poses a challenge. Electromagnetic effects 
are thought to be a prime suspect, but details still need an
explanation. To this end, we study a 
three-dimensional structure of oblique magnetic fields near 
a rotating black hole in vacuum. It has been proposed that such a 
set-up can lead to efficient acceleration when plasma is injected
near a magnetic null point. We focus our attention 
especially on the magnetic field in the immediate neighborhood 
of the magnetic null point, which was previously shown 
to occur in the equatorial plane. By employing the
Line-Integral-Convolution (LIC) method, we visualize the magnetic 
field lines and explore the electric 
lines rising out of the equatorial plane. 

We show the magnetic field structure near the boundary 
of ergosphere, depending on the spin of 
the black hole. Electric field develops a non-vanishing 
component passing through the magnetic null point and
ensuring efficient acceleration of charged particles from 
this particular location near horizon. 
We also examine the effect of translatory boost on the field 
lines. Similarly to the frame-dragging by rotation, the 
linear motion carries field lines along with the black hole.
Position of the magnetic null point recedes from the black
hole horizon as the spin parameter increases. For the extreme
value of $a=1$ the null point can occur outside the ergosphere. 
\end{abstract}
%\maketitle

\section{Introduction}
Magnetosphere is a region surrounding a magnetized body, where 
magnetic effects dominate the motion of electrically charged 
particles. Astrophysical black holes are electrically neutral and do
not support their own magnetic field, nevertheless, they can be 
endowed by a magnetosphere when
embedded in an external magnetic field that is generated by distant
currents. Unlike planetary magnetospheres, black holes bring new 
quality by influencing the structure of electromagnetic fields directly
by their strong gravity \citep{bbr84,p08}. 

Near a rotating black hole, shearing motions are very intense, and 
capable of bending and folding magnetic lines frozen into plasma. 
This process is known to create conditions suitable for magnetic 
reconnection \citep{ka08}. But here we conceive a different setting: 
a magnetically dominated region, where large-scale magnetic 
fields get distorted by purely gravitational effects caused by 
proximity of the black hole. As a result of such an interaction, 
a complicated structure arises of narrow magnetic layers and 
associated neutral points \citep{kk09}. Reconnection can then 
occur when some amount of plasma is brought in the region, 
e.g.\ by the vacuum breakdown and the resulting production
of electron-positron pairs.

Recently, \citet{l11} investigated the role of motion of a Schwarzschild
black hole through an ordered magnetic field. The linear velocity 
component leads to conditions permitting 
the energy extraction by an induced electrical field in the 
direction parallel to the magnetic lines. The released energy
is eventually emitted in the form of an electromagnetic 
signal. However, the motion of a magnetized non-rotating
black hole does not seem to produce magnetic neutral points. 
The formation of such a magnetic topology is an important aspect 
in the context of magnetic reconnection, which typically 
takes place where topologically distinct regions approach each other 
and the magnetic field lines change their connectivity \citep[e.g.,][]{pf00}.

Various processes can lead to the magnetic topology exhibiting 
null points, in particular, these can emerge by complex motions
of the plasma. In this respect the black hole rotation 
brings a new situation, as magnetic field lines become twisted in a 
highly curved spacetime of a rotating black hole, approximately
where the ergosphere forms. This is caused by  
gravito-magnetic interaction of general relativity \citep{i85}. Here, we
examine the resulting structure of the magnetic field, namely, the 
emergence of critical points in a local frame of a physical observer,
resembling the occurrence of X-points, which were studied, e.g.\
recently in the context of special-relativistic reconnection \citep{ko11}. 

In the previous paper \citep{kk09}, we considered a special case
of a uniform magnetic field in perpendicular orientation with respect 
to the black hole spin axis, and we demonstrated that magnetic 
null points can indeed form near a rotating (Kerr) black hole. 
Here, we identify precise location of the null points and we plot
their radial position as a function of spin. The null point emerges, typically,
for $a\gtrsim 0.2$ and it moves gradually away from the horizon until
it crosses the boundary of the ergosphere from an almost extremely rotating
black hole. Furthermore, in this paper the embedded magnetic field is allowed a
general orientation, i.e., it can be inclined in an arbitrary angle with respect
to the rotation axis. Therefore, axial symmetry is broken between the 
gravitational and electromagnetic fields, and this has to result in a truly
three-dimensional structure of magnetic field lines along with 
gravito-magnetically induced electric field 
\citep[see][for a recent review]{rieger11}. In fact it was not obvious whether
the magnetic nulls exist in the oblique case or whether their occurrence is an
artifact of special (perpendicular) orientation of the magnetic field. 
The adopted approach allows us to identify the gravitational effects operating in 
a magnetically dominated system, where a super-equipartition magnetic field
governs the motion of plasma. We indeed find the null points also in a
fully three-dimensional case of an inclined magnetic field.

\section{Aligned and oblique magnetic fields near black hole horizon}
Astrophysical black holes do not support their own intrinsic 
magnetic field; this has to be generated by external currents and 
brought down to horizon by accretion. An active role of accretion 
on transporting large-scale magnetic fields toward horizon has been 
recognized \citep{bl07,rl08}. A black hole can also 
enter a pre-existing magnetic flux tube, and then one asks if the 
process of magnetic reconnection is influenced by strong gravitational 
field near horizon. And does the black hole rotation play a significant 
role in the scenario of this kind?

As we wish to discuss magnetic fields inclined with respect to
the spin axis, and we also want to include the fast 
translatory motion, the following picture appears to be appropriate:
Kerr black hole traversing a ``magnetic filament'', described as an
extended 
(largely one-dimensional) flux tube. Such an idea can be motivated
observationally, by highly ordered and elongated arcs (of 
about $100\mu$G) that are seen in Galactic Center, 
within a few parsecs from Sagittarius 
A* compact radio source \citep[Sgr~A*;][]{ym84}. They are thought to represent 
large-scale magnetic flux tubes that are illuminated by
synchrotron emission from relativistic electrons \citep{ln04,m06,f10}.

Given a limited resolution that can be achieved with current
imaging techniques, the magnetic filaments cannot be traced down to the 
characteristic size of the black hole.\footnote{Gravitational radius
$r_{\rm{g}}=GM/c^2\approx1.48\times10^{13}\,M_8$~cm, where the central
black hole mass is expressed in terms of $M_8{\equiv}M/(10^8M_{\odot})$.
The velocity of the Keplerian orbital motion of a particle 
is then $v_{_{\rm{}K}}\approx2.1\times10^{10}
(r/r_{\rm{}g})^{-1/2}{\rm{cm\;s}}^{-1}$. The corresponding orbital
period is $T_{_{\rm{}K}}\approx3.1\times10^3(r/r_{\rm{}g})^{3/2}M_8$~s.
Hereafter, we use a dimensionless form of geometrized units,
where all quantities are scaled with the black hole mass;
$M$ does not appear in the equations explicitly.} Therefore,
the actual mapping of the magnetic structures near black holes 
is not directly possible (this may change with new interferometrical
techniques in near future). The role of magnetic structures in accelerating the 
particles is suggestive, especially because they could help us 
to understand the origin of particle acceleration and the resulting 
signatures in the electromagnetic signal.

Naturally, in astrophysically 
realistic solutions the role of non-ideal plasma will need be included. Currently, 
neither of the frequently discussed limits (i.e., vacuum vs.\ force-free
approximations) are able to account for both the plasma currents 
as well as the accelerating electric fields. This task will require
resistive MHD, which is, however, beyond the scope of this brief 
paper.

\begin{figure}[tbh!]
\centering
\includegraphics[width=0.32\textwidth]{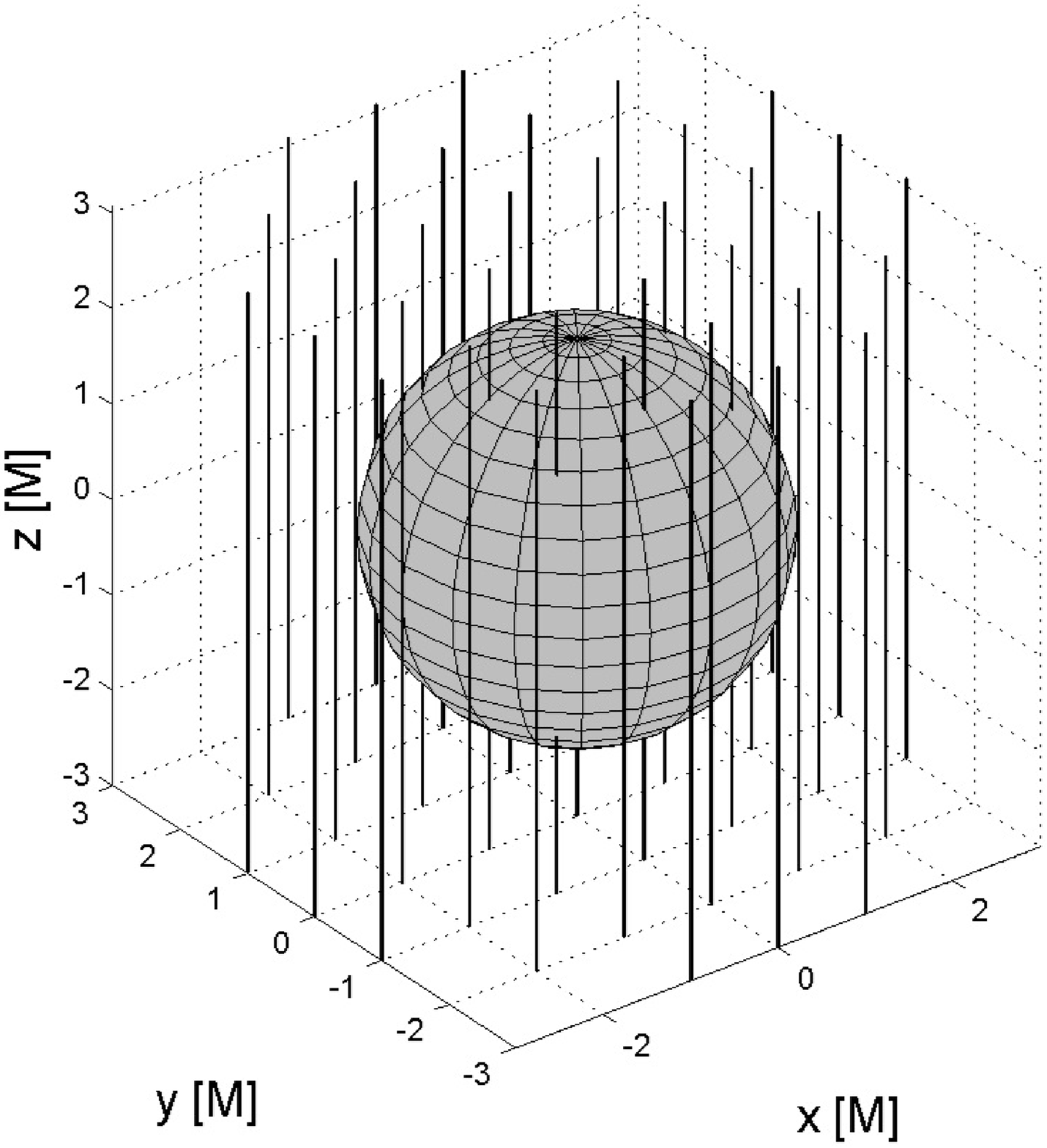}%{mag_a0_Bx0_Bz1.eps}
\hfill
\includegraphics[width=0.32\textwidth]{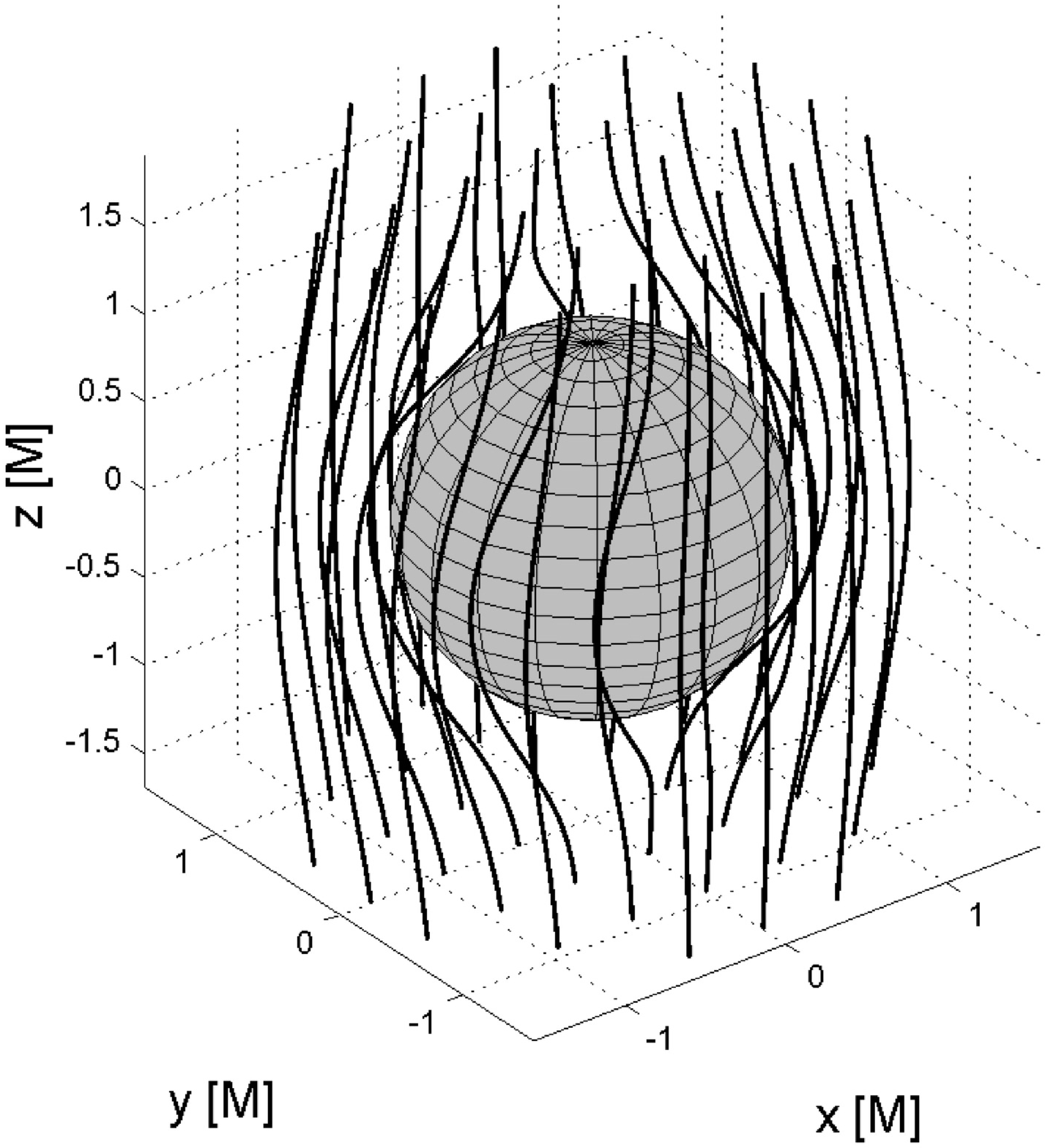}%{mag_a1_Bx0_Bz1.eps}
\hfill
\includegraphics[scale=0.32, clip, trim= 50mm 5mm 50mm 15mmh]{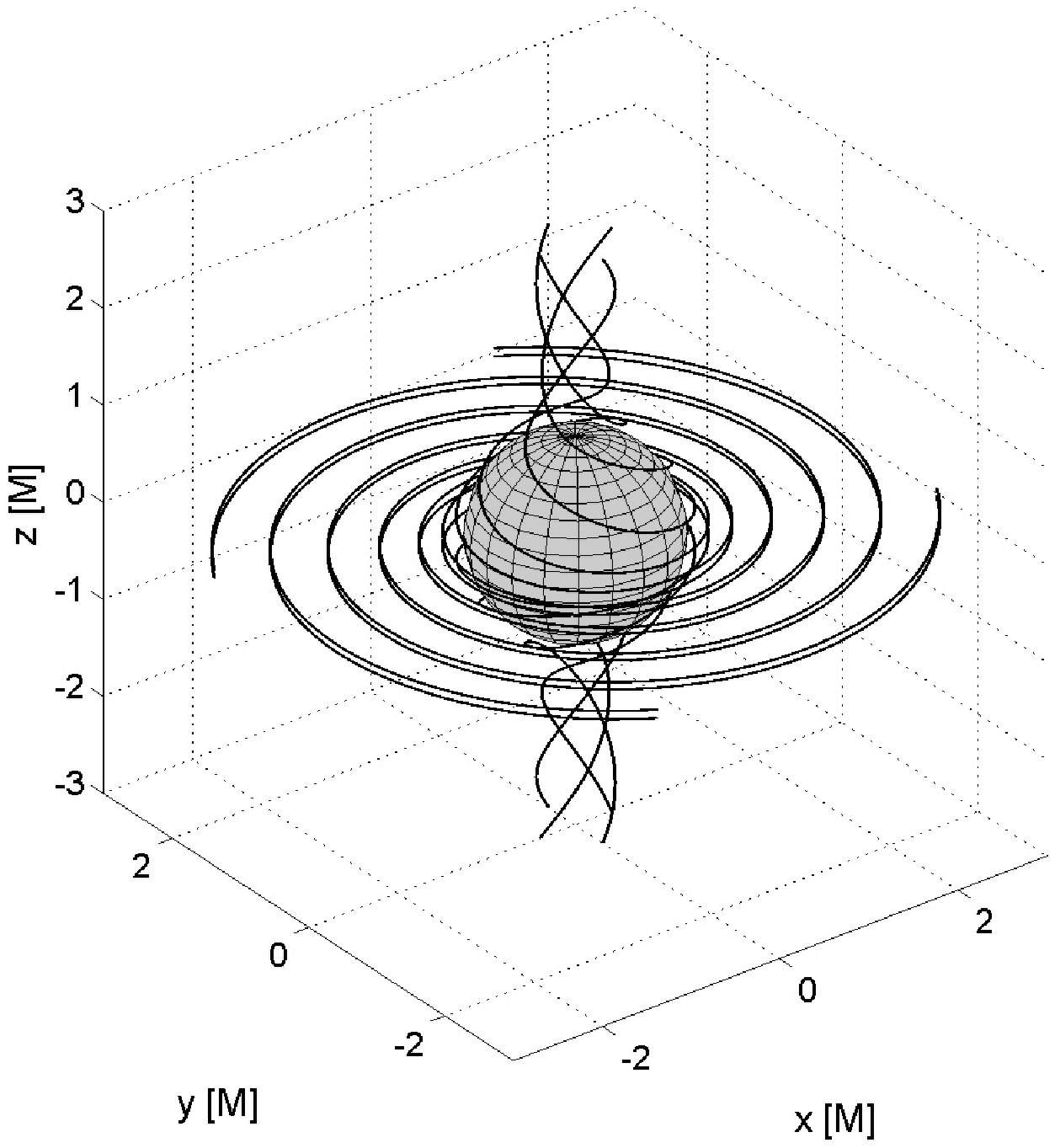}%{el_meissner_FFOFI_3d.eps}
\caption{Asymptotically uniform magnetic field aligned with
rotation axis of the black hole in vacuum. Tetrad components 
are plotted with respect to frame of free-falling observers in stereometric 
projection. Left: magnetic lines near a non-rotating black hole appear to
be perfectly homogeneous (in these suitably defined coordinates). 
Middle: magnetic field aligned with the rotation axis of an maximally 
rotating ($a=1$) Kerr black hole, when they are expelled out of
the horizon. Right: electric lines corresponding to 
the magnetic field shown in the middle panel. A toroidal component is 
induced by the gravito-magnetic action of the rotating black hole
\citep[see][for more details]{k11}.}
\label{fig1}
\end{figure}

\subsection{The effect of black hole rotation and translatory motion}
Starting from  \citet{kl75} and \citet{bd76}, the organised 
electromagnetic test fields surrounding 
black holes have been discussed and their astrophysical consequences
considered in various papers. The case of oblique 
geometry, however, has been explored only partially \citep{bj80}. 
This is caused especially by the fact that the off-equatorial fields
are lacking any symmetry, and so are more difficult to visualise. 
Also, qualitatively new effects on the field 
structure were not expected. 

Nevertheless, in \citet{kk09} we were 
able to demonstrate that the asymptotically perpendicular field 
lines develop a magnetic neutral point in the equatorial plane. 
This is interesting because such structures of the magnetic field
are relevant for processes of electromagnetic acceleration. 
The magnetic null point emerges in a local 
physical frame, and could trigger reconnection. 
However, the asymptotically perpendicular field is a rather 
exceptional case. Therefore, here we investigate near-horizon 
magnetic structures in more detail, also for a general 
orientation of the magnetic field.

The main objective of this discussion is to track the location of
magnetic null points and to explore a complex three-dimensional
configuration inside the ergosphere. We consider the form of magnetic 
lines together with the induced electric lines for different 
values of the model parameters: the inclination angle of the asymptotic 
magnetic field $\theta_{\rm{}o}$ ($=\arctan(B_{\perp}/B_{\parallel}$), 
the black hole spin $a$ ($a^2\leq M$), and the boost velocity $\beta$ 
($\beta^2\equiv v_x^2+v_y^2+v_z^2<1$). We observe the layers of 
alternating magnetic orientation to occur also in the general case, i.e., 
when the black hole rotates and moves with respect to an oblique
magnetic field. However, the three-dimensional structure of the field
lines is very complicated as they become highly entangled around the 
null point. 

\begin{figure}[tbh!]
\centering
\includegraphics[scale=0.36, clip,trim=30mm 5mm 65mm 5mm]{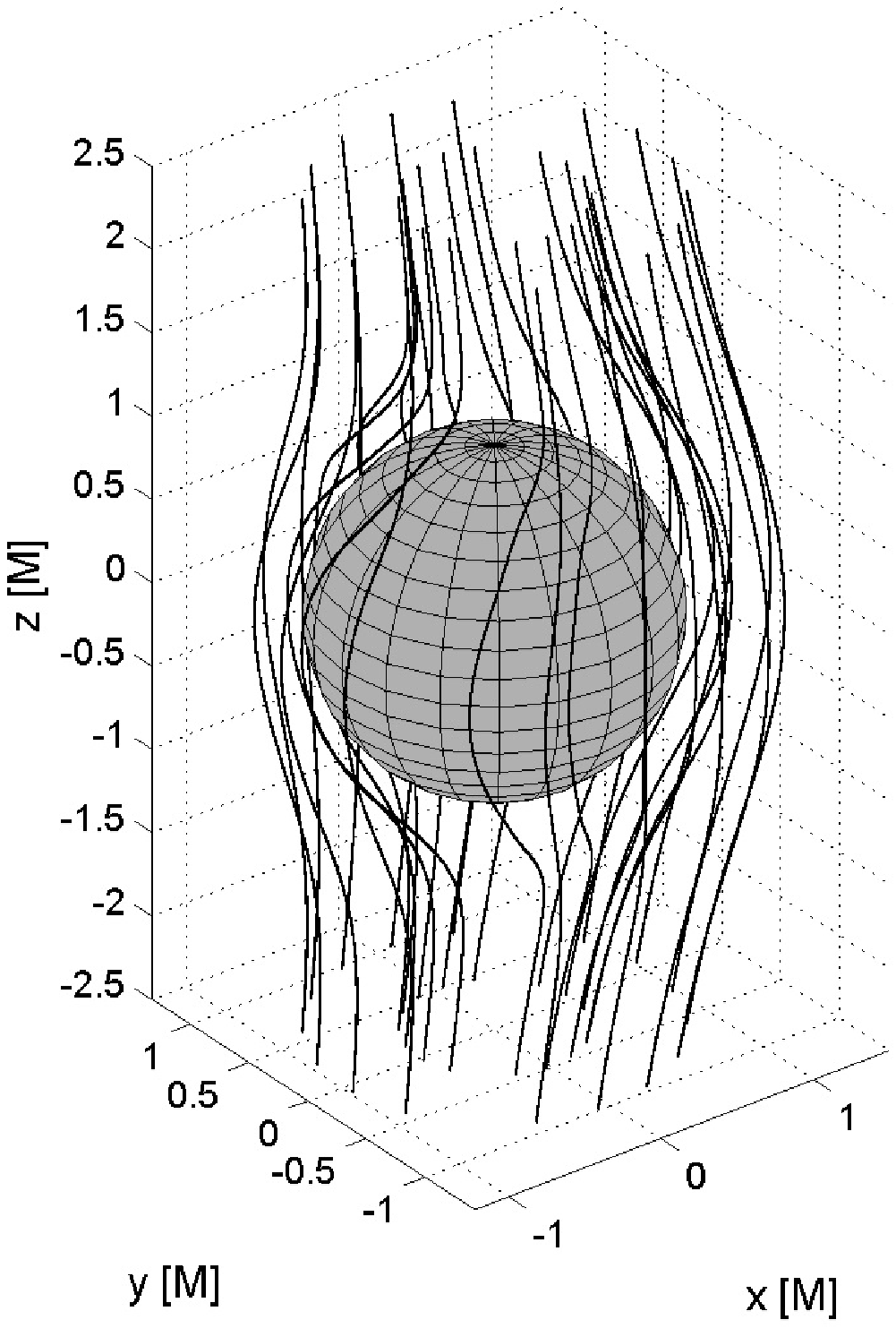}%{drift_3d_FFOFI_vx0.1_vy-0.1.eps}
\hfill
\includegraphics[scale=0.36, clip,trim=30mm 5mm 65mm 5mm]{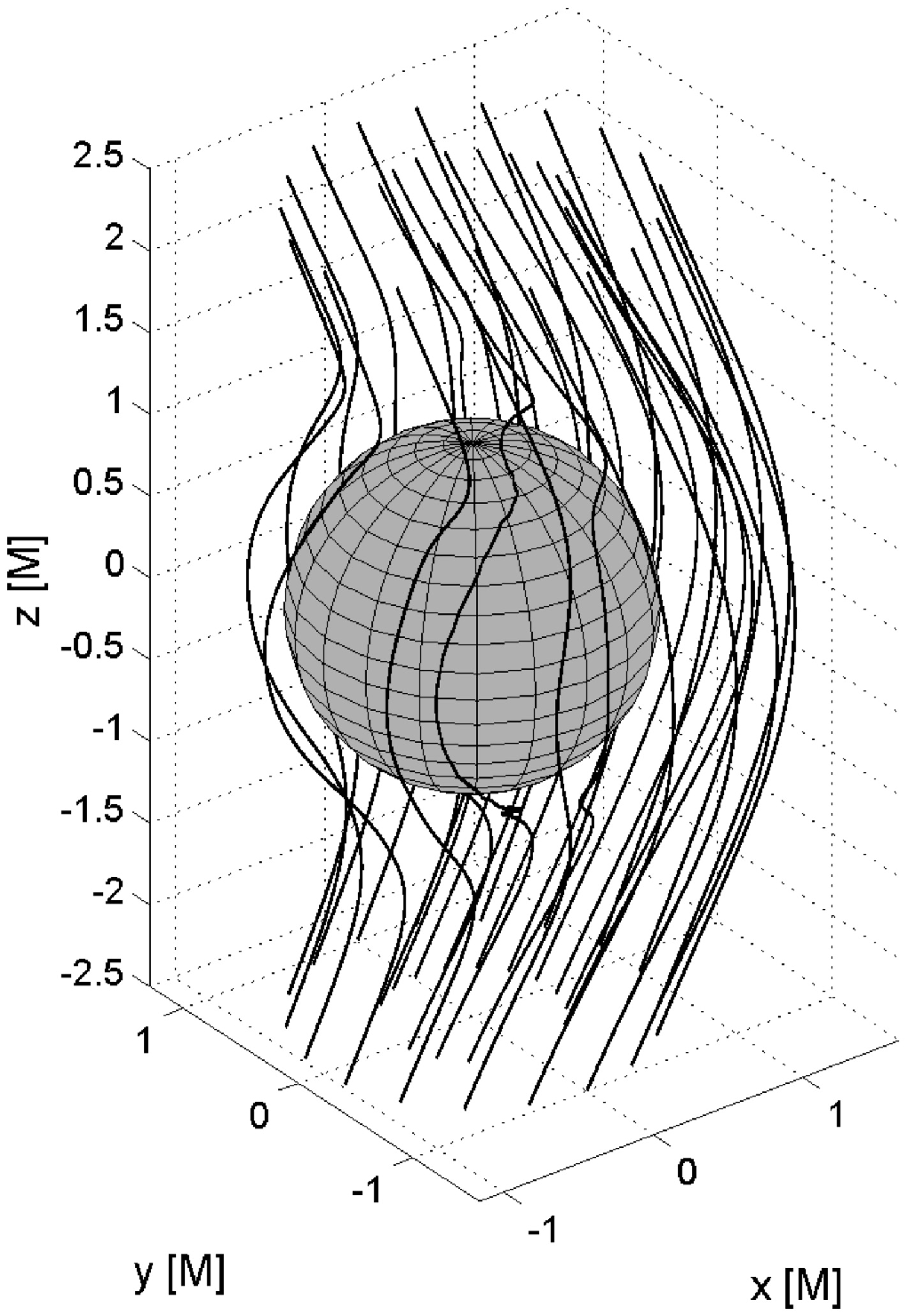}%{drift_3d_FFOFI_vx0.3_vy-0.3_enhanced.eps}
\hfill
\includegraphics[scale=0.36, clip,trim=30mm 5mm 65mm 5mm]{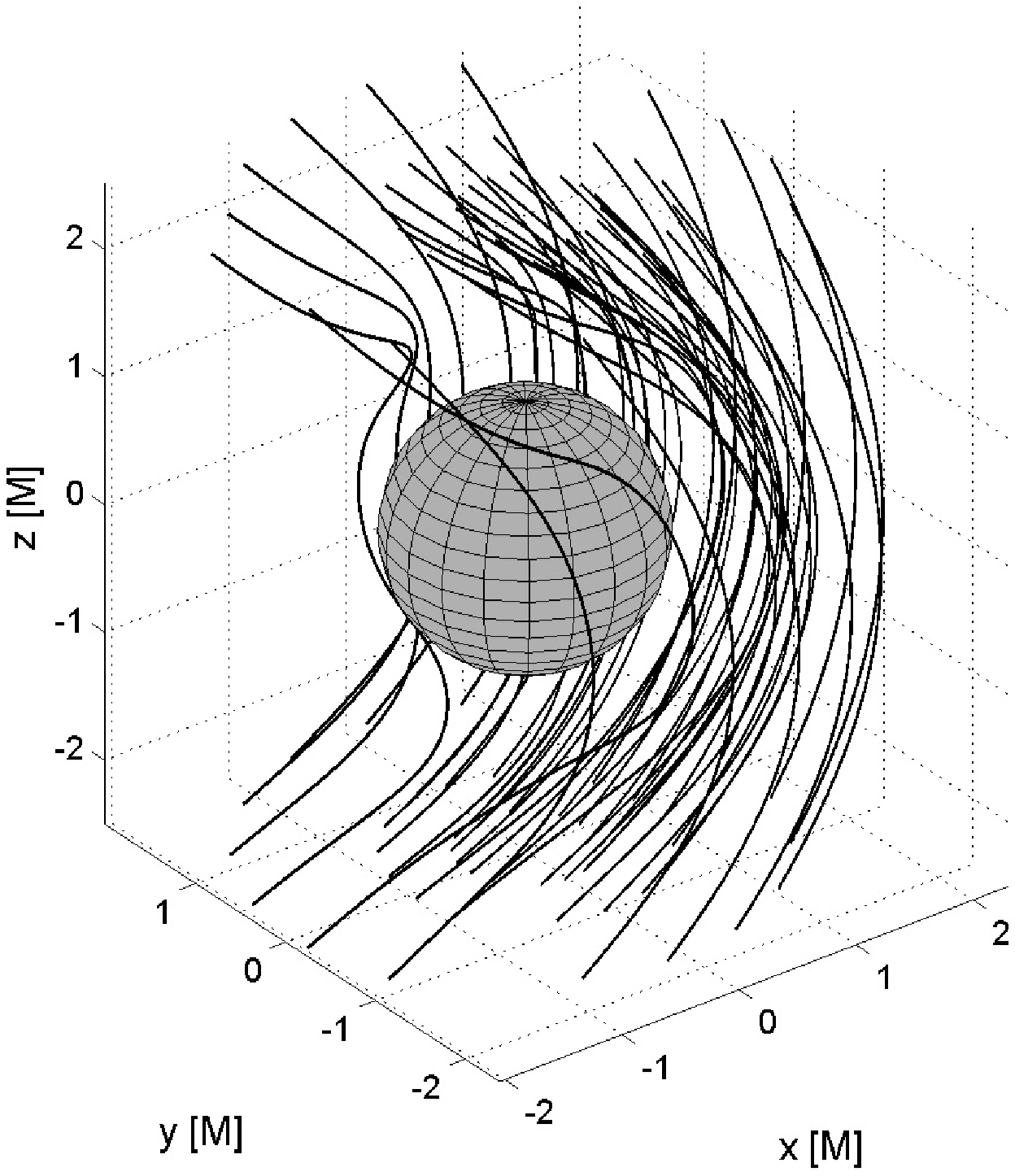}%{drift_3d_FFOFI_vx0.7_vy-0.7.eps}
\caption{The effect of black hole motion on magnetic lines from the previous
figure. The case of extreme spin, $a=1$, with a gradually growing velocity of the
translatory boost along $x$-direction: $v_x=0.1$ (left),  $v_x=0.3$ (middle), 
and $v_x=0.7$ (right).}
\label{fig2}
\end{figure}

We specify the gravitational field by Kerr metric \citep{mtw73}.
Our starting form of the 
electromagnetic field is a stationary solution of Maxwell's test-field 
electro-vacuum equations in the given curved spacetime.\footnote{These are 
astrophysically acceptable approximations which reflect the fact that 
black holes can only acquire a negligibly small electric charge, while 
cosmic electromagnetic fields are not strong enough to influence the 
spacetime metric significantly.} The electromagnetic four-potential $\bm{A}$ 
can be then written as superposition of two contributions: 
$\bm{A}=B_{\parallel}\,\bm{a}_{\parallel}+B_{\perp}\,\bm{a}_{\perp}$,
where $B_{\parallel}$ and $B_{\perp}$ define the magnitudes of the two
parts.

The aligned field has two non-vanishing components of the normalized 
electromagnetic four-potential,
\begin{eqnarray}
a_{t\parallel} &=& B_{\parallel}a\Big[r\Sigma^{-1}\left(1+\mu^2\right) -1\Big], \label{mf1}\\
a_{\phi\parallel} &=& B_{\parallel}\Big[{\textstyle\frac{1}{2}}\big(r^2+a^2\big)
 -a^2r\Sigma^{-1}\big(1+\mu^2\big)\Big] \sigma^2 \label{mf2},
\end{eqnarray}
where we use Boyer-Lindquist $t$, $r$, $\theta$, and $\phi$ 
dimension-less spheroidal coordinates ($\mu=\cos\theta$, $\sigma=\sin\theta$). 
Eqs.\ (\ref{mf1})--(\ref{mf2}) represent an asymptotically homogeneous 
magnetic field \citep{w74}. 

On the other hand, the perpendicular to axis component of the field is 
given by \citep{bj80}
\begin{eqnarray}
a_{t\perp} &=& B_{\perp}a\Sigma^{-1}\Psi\sigma\mu, \label{mf3} \\
a_{r\perp} &=& -B_{\perp}(r-1)\sigma\mu\sin\psi, \\
a_{\theta\perp} &=& -B_{\perp}\Big[\big(r\sigma^2+\mu^2\big) a\cos\psi 
 + \Big(r^2\mu^2+\big(a^2-r\big)(\mu^2-\sigma^2)\Big) \sin\psi\Big], \\
a_{\phi\perp} &=& -B_{\perp}\Big[\Delta\cos\psi+\big(r^2+a^2\big)
 \Sigma^{-1}\Psi\Big] \sigma\mu, \label{mf4}
\end{eqnarray}
with $\Sigma(r,\mu)$ and $\Delta(r)$ being the Kerr metric functions,
$\psi\equiv\phi+a\delta^{-1}\ln\left[\left(r-r_+\right)/\left(r-r_-\right)\right]$, 
$\Psi=r\cos\psi-a\sin\psi$, $\delta=r_+-r_-$, and $r_{\pm}=1\pm\sqrt{1-a^2}$.
Knowing the complete set of four-potential components the magnetic field structure
is fully determined: $F_{\mu\nu}\equiv A_{[\mu,\nu]}$.

\begin{figure}[tbh!]
\centering
\includegraphics[width=0.49\textwidth]{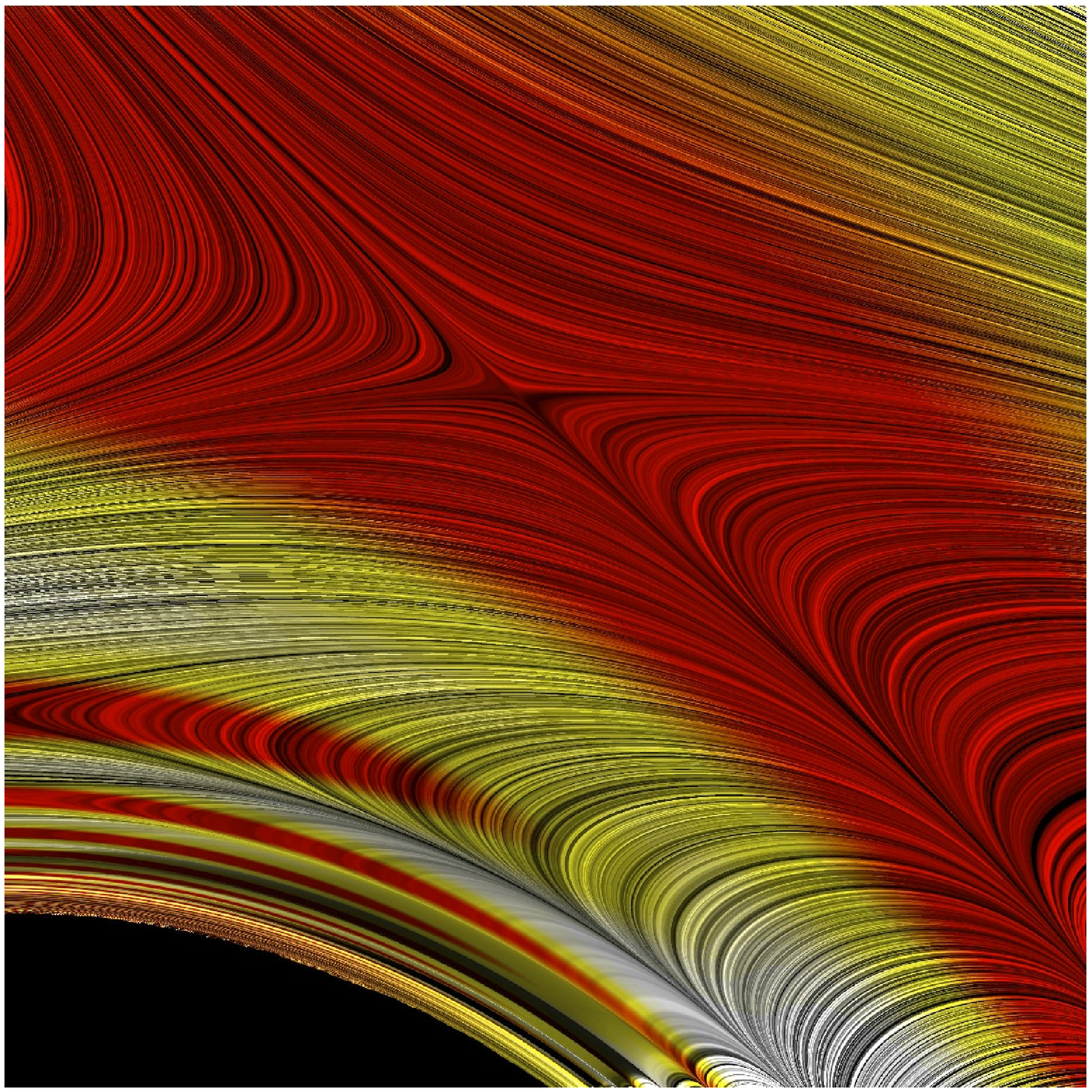}%{bhmag7.eps}
\hfill
\includegraphics[width=0.49\textwidth]{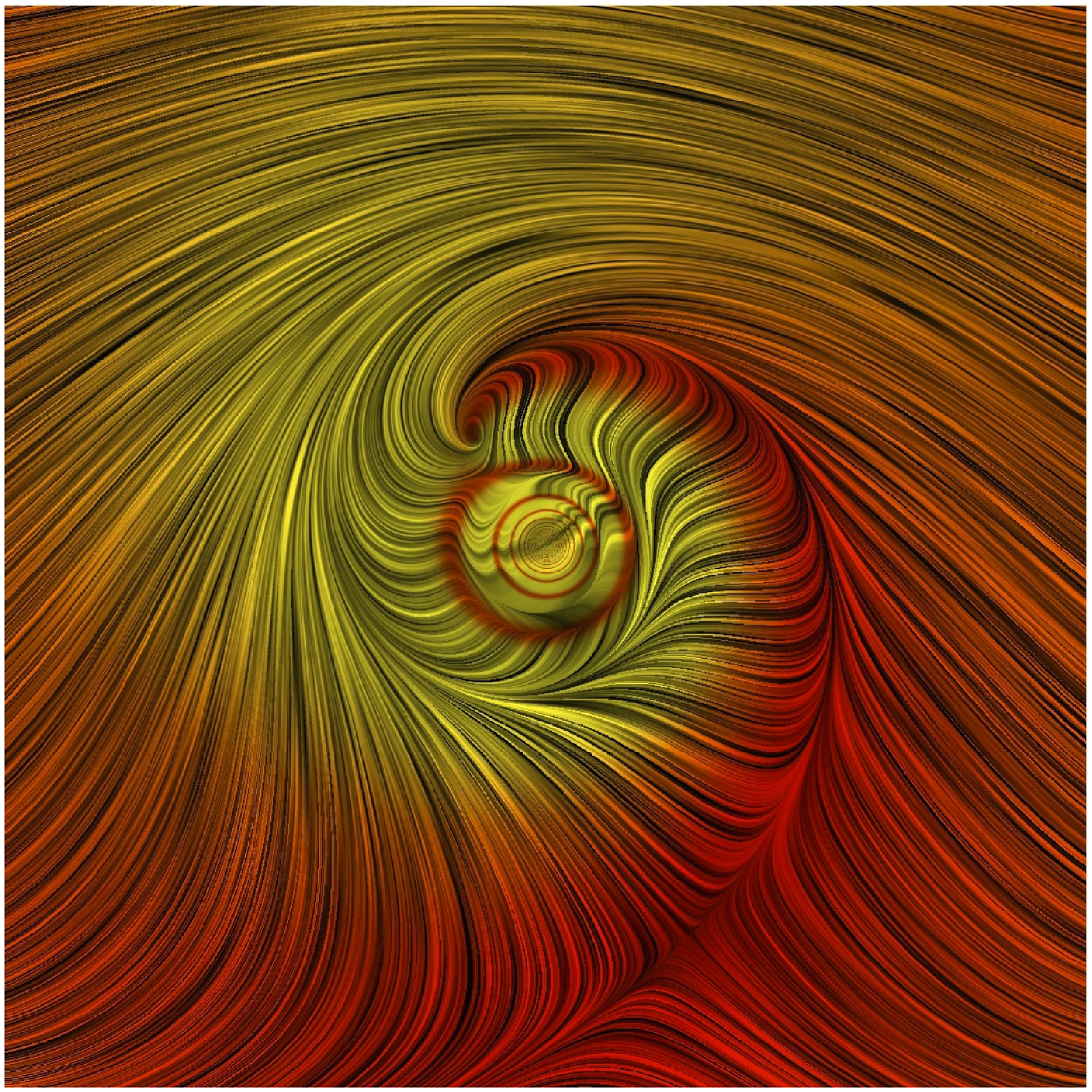}%{bhmag6.eps}
\caption{Perpendicular magnetic field in the equatorial plane 
$(x,y)\equiv(r\cos\phi,r\sin\phi)$, i.e., viewed along the rotation 
axis of an extreme ($a=1$) black hole. While the field is steady and 
uniform far from the black hole, its structure becomes highly 
entangled near horizon. Left panel: the neighborhood of the 
magnetic null point, which reveals itself clearly in the upper part of 
the plot. The black circular section in the bottom left corner denotes 
the horizon, $r=r_+(a)$. Right panel: the layered 
structure of the magnetic field close above the horizon is revealed 
by using the radial coordinate $\xi(r)\leq0.6$ (i.e., $r_+< r\leq2.5GM/c^2$;
$r=r_+$ collapses to the center of the plot). The magnetic
null is now exhibited in the lower part of the plot.
The colour scale indicates the magnetic intensity (in arbitrary 
units), ranging from vanishingly small values (red) across the moderate 
(yellow) to the maximum (white).}
\label{fig3}
\end{figure}

Equations (\ref{mf3})--(\ref{mf4}) describe the field lacking the axial 
symmetry. Such a situation cannot be strictly stationary, however, 
the alignment time-scale is very long, and so we can neglect the 
associated energy losses here \citep{kl03}. The field line structure 
depends also on the motion of observers determining the field 
components. In order to construct and discuss our plots,
we employ a free-falling physical frame, evaluate the 
electromagnetic tensor with respect to the local frame, and use 
these components to draw the field lines.

\begin{figure}[tbh!]
\centering
\includegraphics[width=0.37\textwidth]{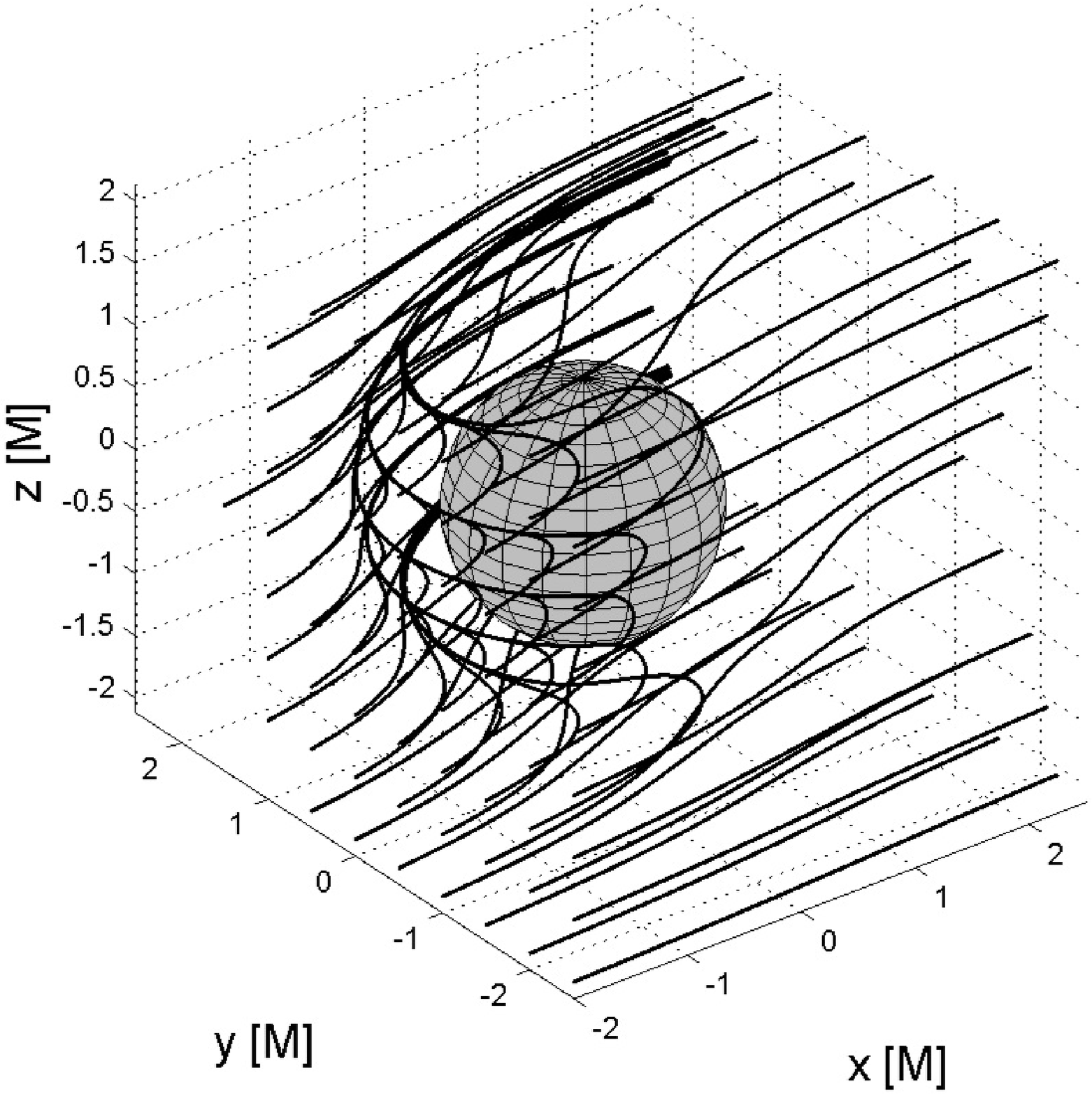}%{mag_3D_Bx1_Bz0_vz_0c.eps}
\hfill
\includegraphics[width=0.25\textwidth]{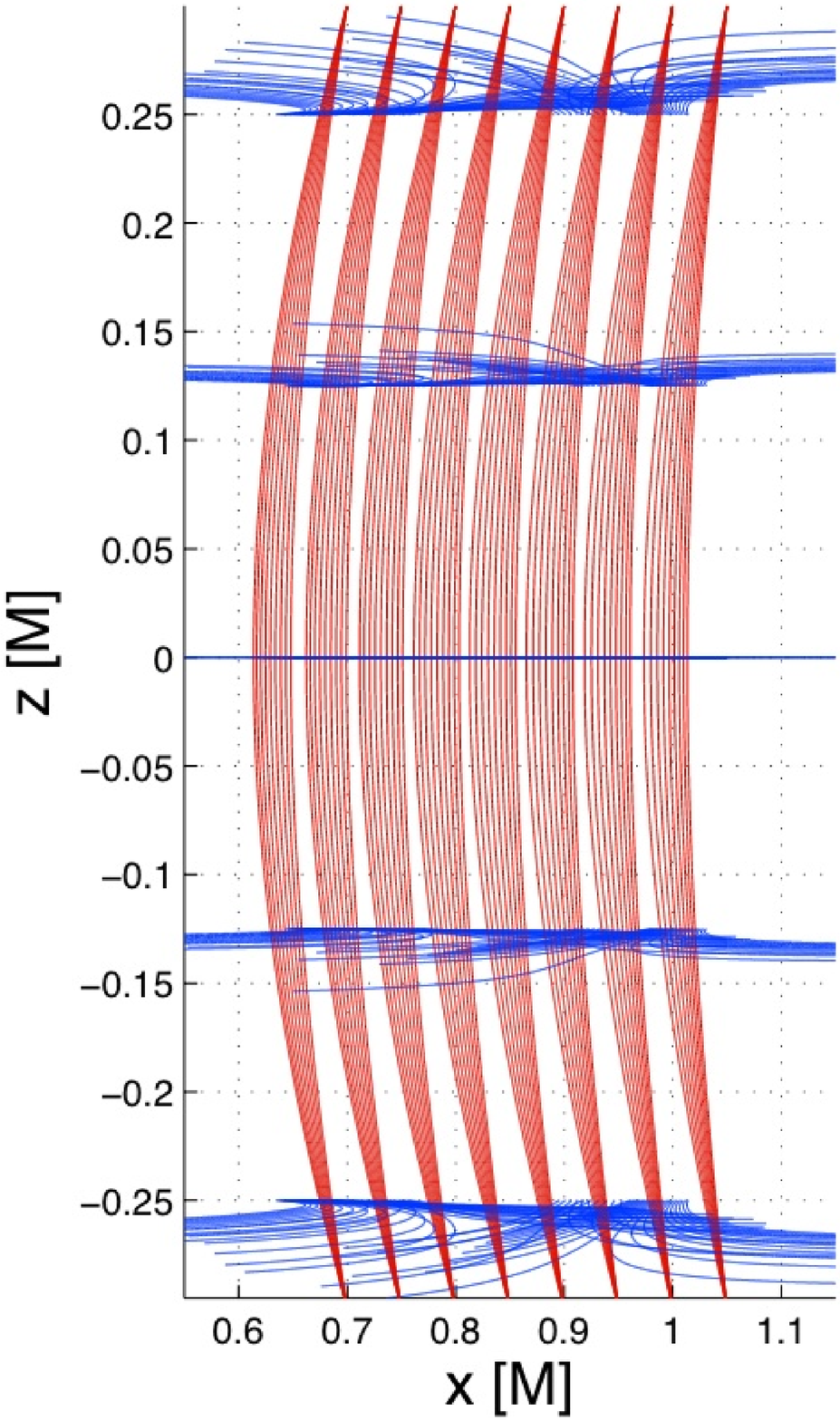}%{NL_3D_new_fo.eps}
\hfill
\includegraphics[width=0.36\textwidth]{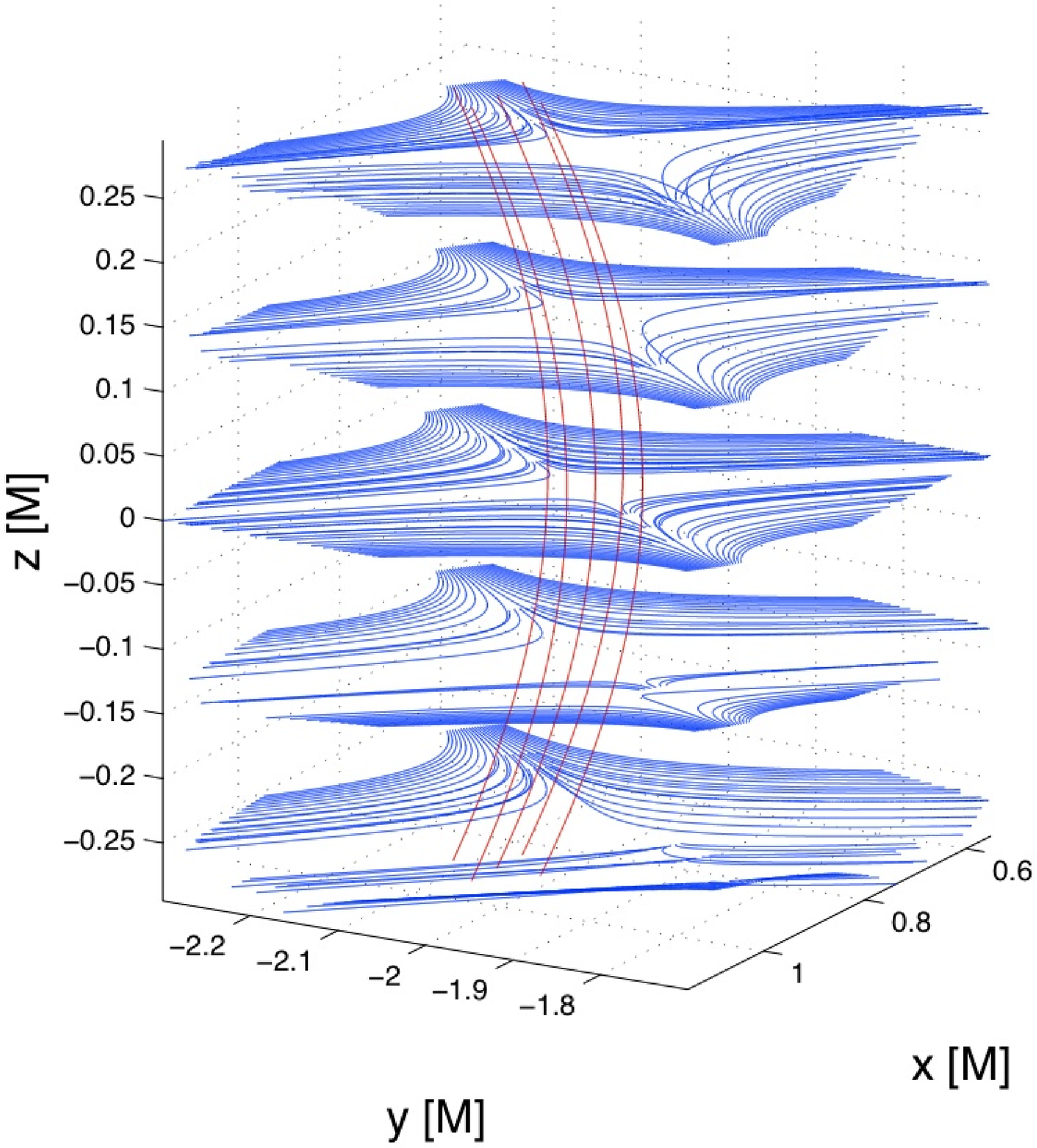}%{NL_3D_more_el2.eps}
\caption{Left panel: magnetic field asymptotically perpendicular to the 
rotation axis ($z$) of an $a=1$ black hole, centered on the origin. 
Clearly seen is the effect of dragging by rotation of the black hole.
Middle: side-view projection of the equatorial ($z=0$) plane in which 
magnetic field lines (blue) reside, while the induced electric lines (red) are 
rising out of the plane. Outside the equatorial plane, also the magnetic field 
acquires some non-zero vertical component allowing the particle acceleration. 
Right: a three-dimensional detail around the equatorial plane reveals a 
magnetic null point. Again, the electric lines are shown to pass through 
the null point in the vertical direction.}
\label{fig4}
\end{figure}

In absence of perpendicular 
component ($B_{\perp}=0$), the field is relatively uncomplicated 
(see figure \ref{fig1}). Although the frame-dragging acts also on the 
aligned field lines, their shape can be integrated to find the 
surfaces of constant magnetic flux in a closed (analytical) form.
Previously, the aligned fields were explored especially in context 
of magnetic field expulsion from a maximally rotating black hole 
\citep[and how this is changed when a conducting medium surrounds 
the black hole; cf.][]{km07}. In our notation, the example in Fig.\ \ref{fig1}
refers to zero velocity of the translatory boost, i.e.\ $\beta=0$. 
Once we include a nonzero boost velocity, 
the structure of the aligned field becomes more complicated 
(figure \ref{fig2}). This is mainly due to an interplay
between the boost of the black hole and its rotation acting jointly 
on the (originally) aligned field. As $\beta$ increases, the magnetic 
lines are progressively puffed out of horizon and wound around it
(see \citeauthor{k11} \citeyear{k11}, for more examples and details).

\subsection{Neutral points of the magnetic field}
Let us now explore the case of magnetic field with a non-vanishing 
component inclined with respect to rotation axis ($B_{\perp}\neq0$). 
In fact, \citet{kk09} explored a strictly perpendicular case. 
Confining the magnetic lines in the equatorial plane 
$\theta=\pi/2$, the nested structure of magnetic layers 
emerges. These are essential for the magnetic 
reconnection. 

Near-horizon structure of magnetic lines is visualized in 
figure~\ref{fig3} by using the Line-Integral-Convolution 
(LIC) method in Matlab. This technique allows us to identify
clearly the location of neutral points. It turns out to be
particularly useful here with the general orientation of
the asymptotic field direction, as the global solution for the field
lines is too cumbersome. Further,
by introducing $\xi(r;a)\equiv1-r_+(a)/r$ as a new radial coordinate, 
the horizon surface $r=r_+$ collapses into the center and the 
layered structure of magnetic sheets is seen in more detail.

\begin{figure}[tbh!]
\centering
\includegraphics[width=0.75\textwidth]{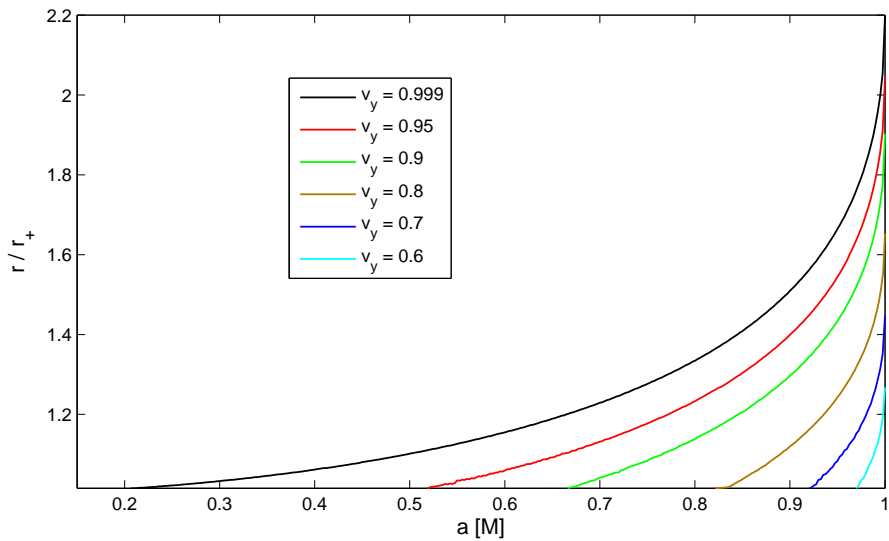}%{findnull2.eps}
\caption{Position in the equatorial plane of the magnetic null point 
corresponding to a magnetic field asymptotically perpendicular  with respect to 
rotation axis. Fast translatory motion and fast rotation are both important 
for the emergence of null points. 
Several curves are shown as a function of the black hole spin $a$ 
for different velocity $v_y$ (i.e., in the direction along $y$-axis), 
which is set in the direction perpendicular with respect to the rotation axis, 
as well as perpendicularly to the asymptotic uniform magnetic field.}
\label{fig5}
\end{figure}

\begin{figure}[tbh!]
\centering
\includegraphics[width=0.49\textwidth]{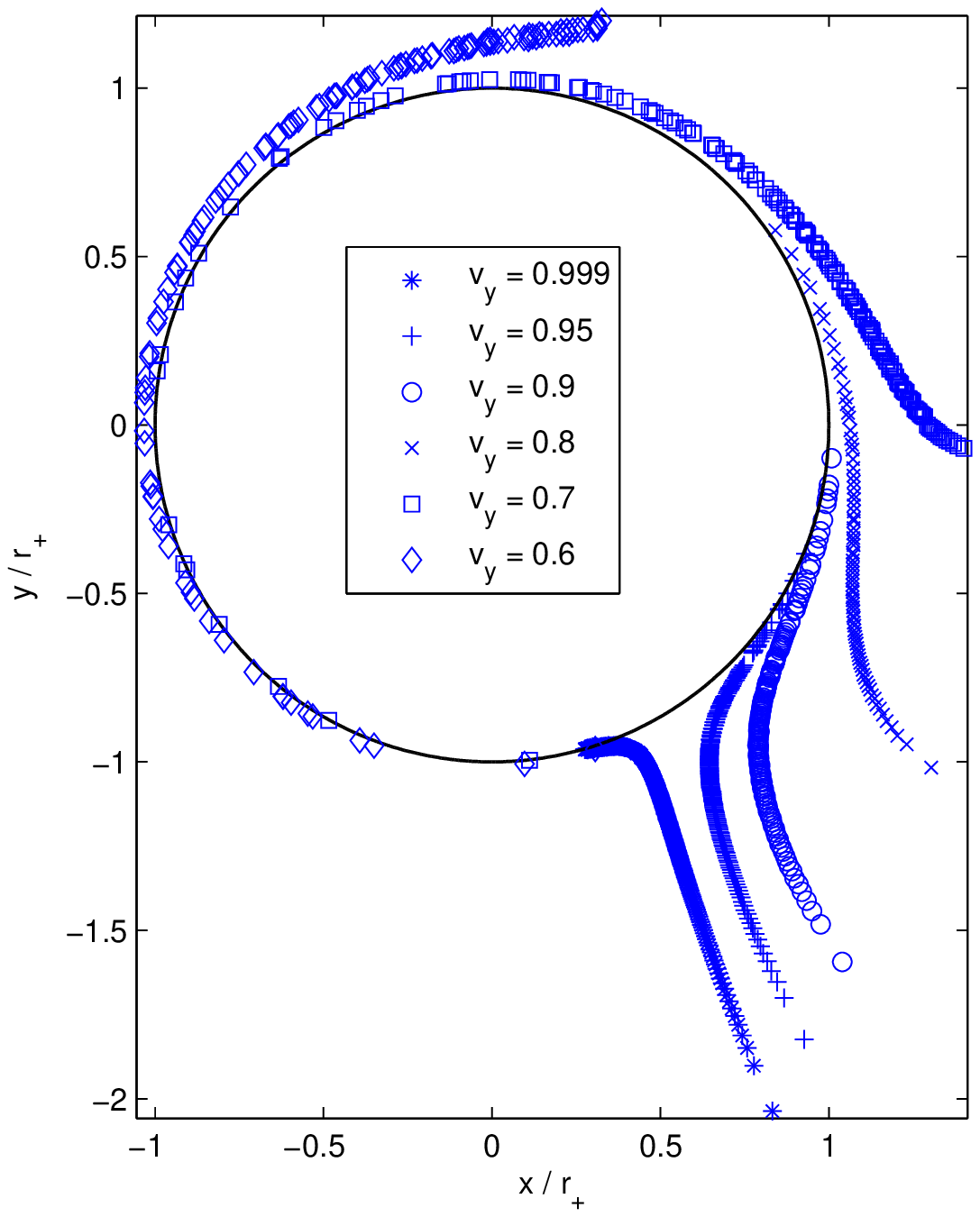}%{NPpoz.eps}
\hfill
\includegraphics[width=0.435\textwidth]{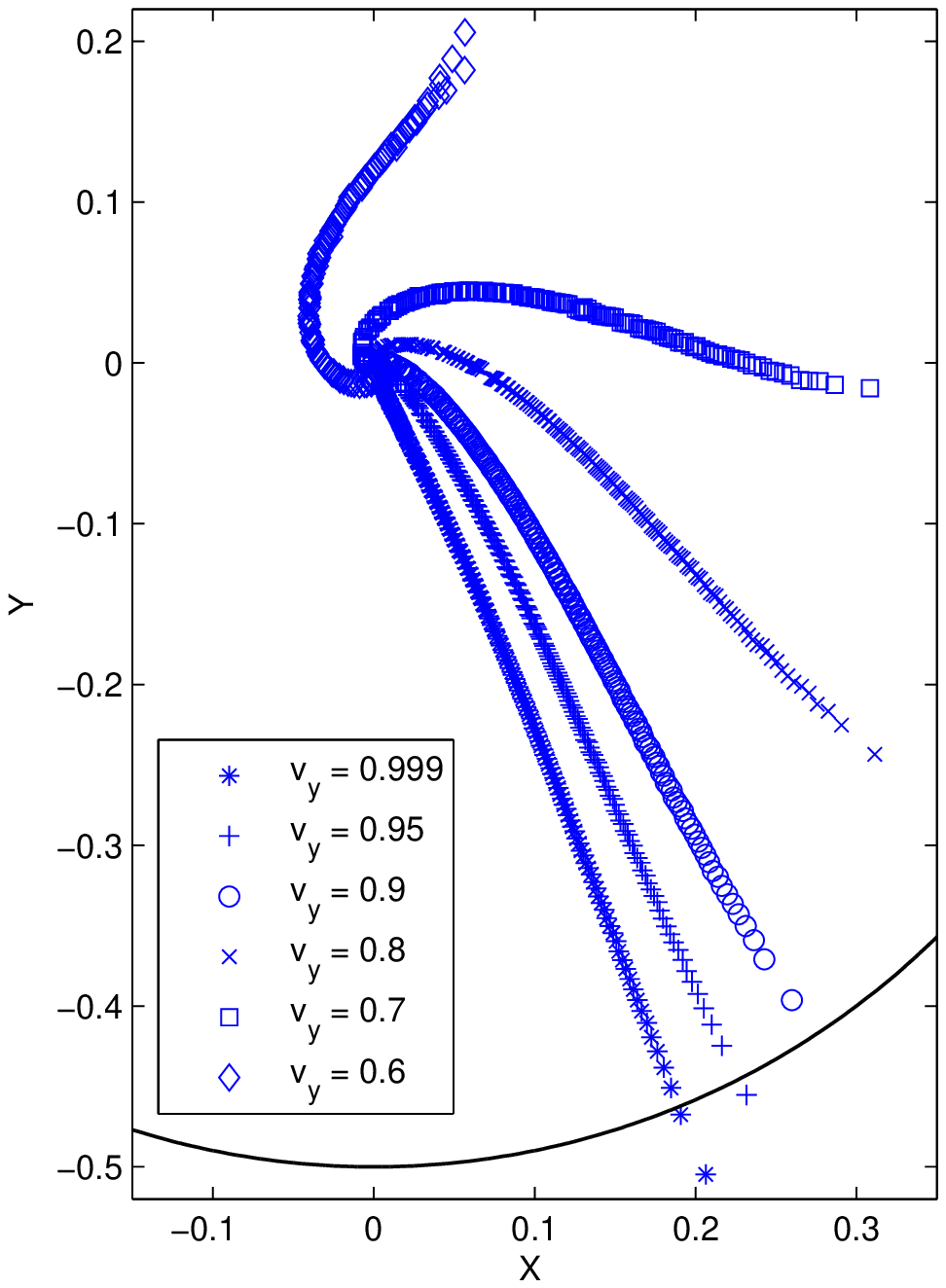}%{NPpoz_tur.eps}
\caption{Position of the magnetic null points which emerge in the equatorial plane 
as the spin and the drift velocity are gradually increased. Different markers indicate the
magnitude of translation velocity as specified in the inset (only $v_y$ component of the velocity
in the equatorial plane is assumed to be non-zero). Left panel: Given the velocity, null points form 
separate traces of a gradually increasing spin $a$: outer endpoints correspond 
to the extreme spin $a=1$; the spin decreases monotonically along each track as one 
approaches the horizon (denoted by the black circle). Right panel: The same as on the left side, 
but now plotted with respect to the rescaled polar radial coordinate $\xi$. This helps us to resolve
more clearly the narrow region just above the horizon (in the origin of the graph). The outer
circle indicates the equatorial radius of the ergosphere of the extreme spin black hole.}
\label{fig6}
\end{figure}

\begin{figure}[tbh!]
\centering 
\includegraphics[width=0.45\textwidth]{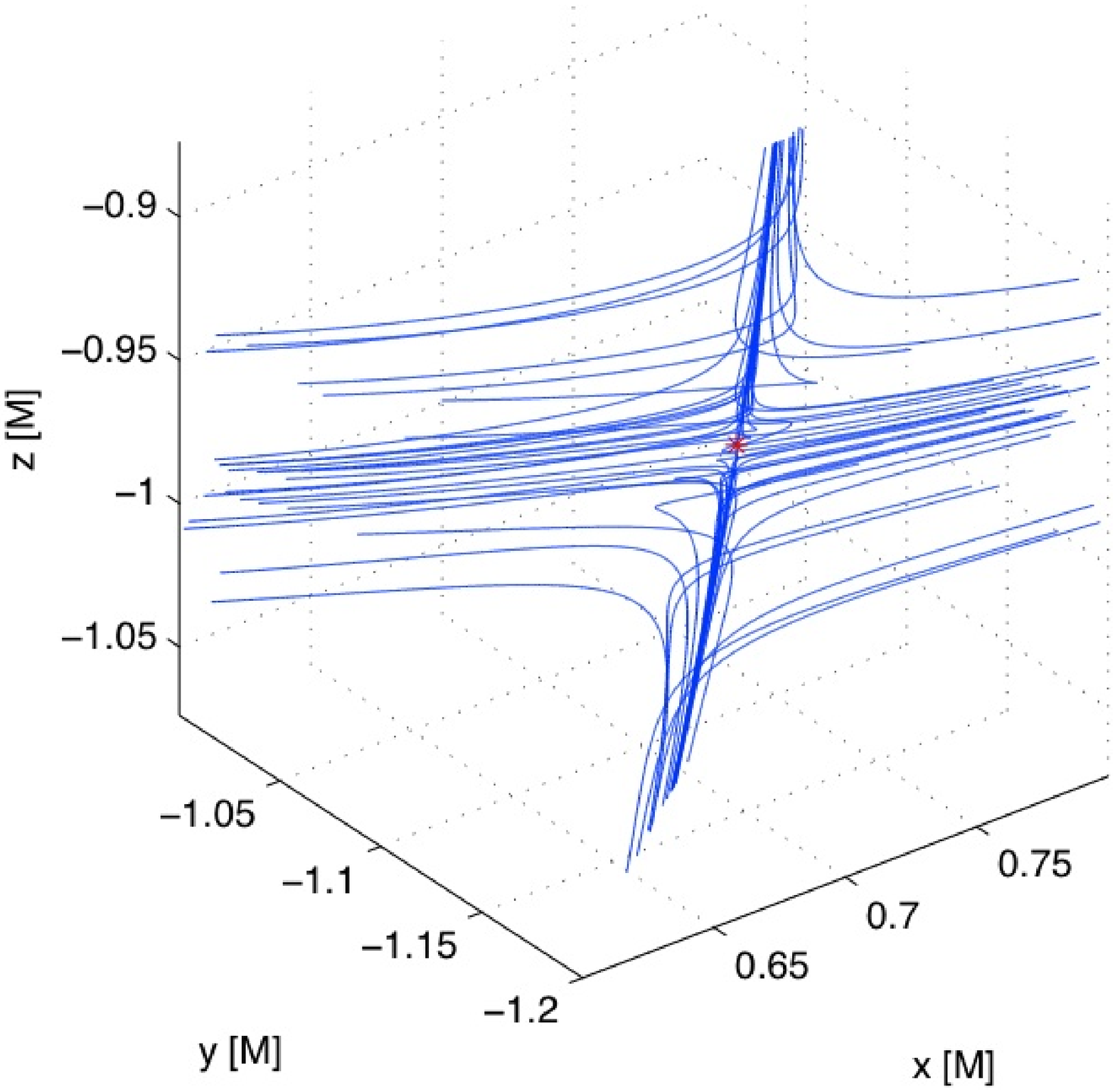}%{3dnull2.eps}
\hfill
\includegraphics[width=0.53\textwidth]{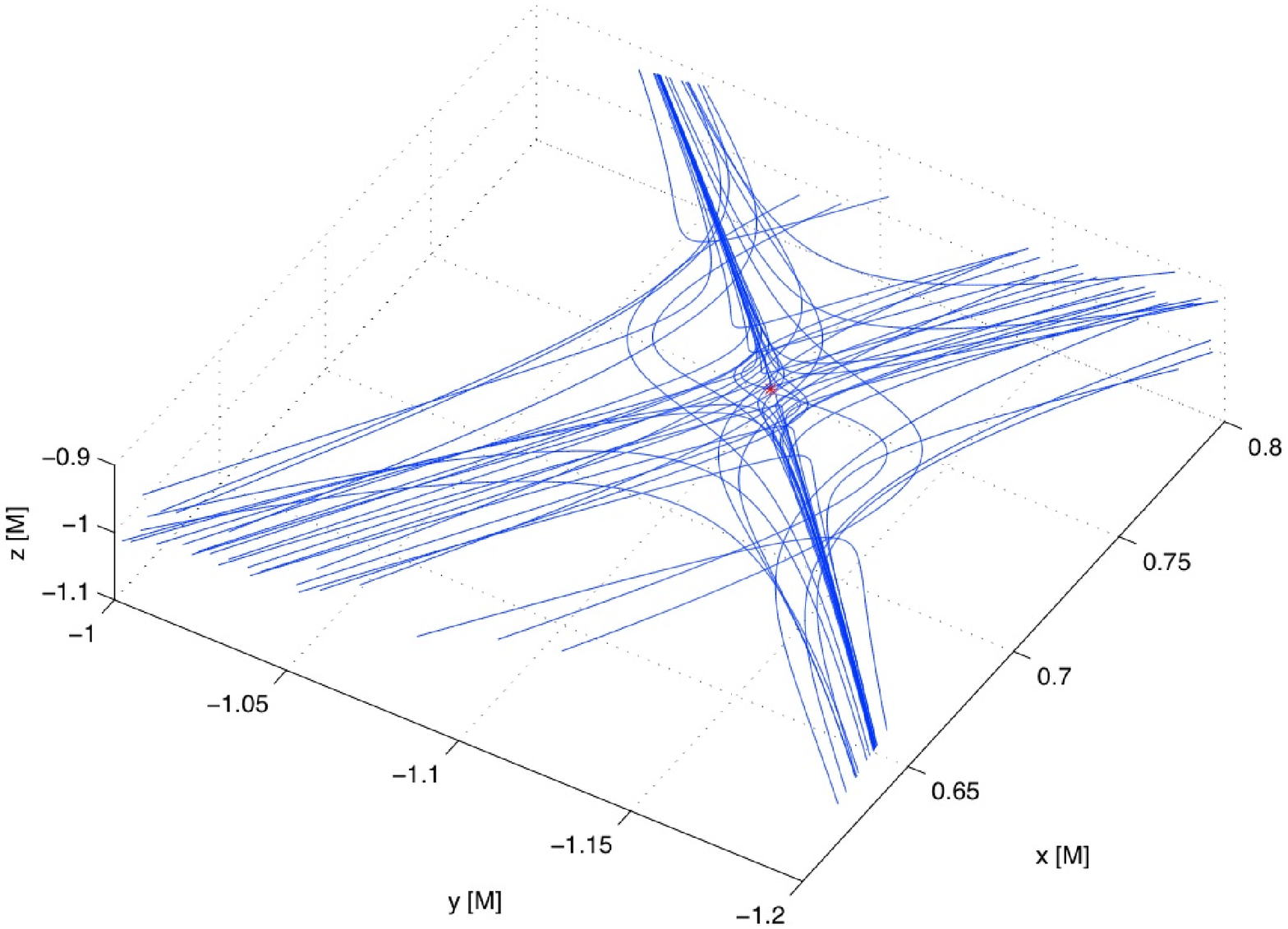}%{3dnull3.eps}
\caption{Structure of the oblique magnetic field near the null point (depicted by the red marker). 
The null point occurs near above the horizon. Here we consider the extremal black hole 
with a rapid drift velocity ($v_x=v_y=0.7$, $v_z=0$) and the background of an inclined 
magnetic field ($B_{x}>0$, $B_{x}=B_{z}$). Two different views of the same system are
shown.}
\label{fig7}
\end{figure}

\begin{figure}[tbh!]
\centering 
\includegraphics[scale=0.85,clip]{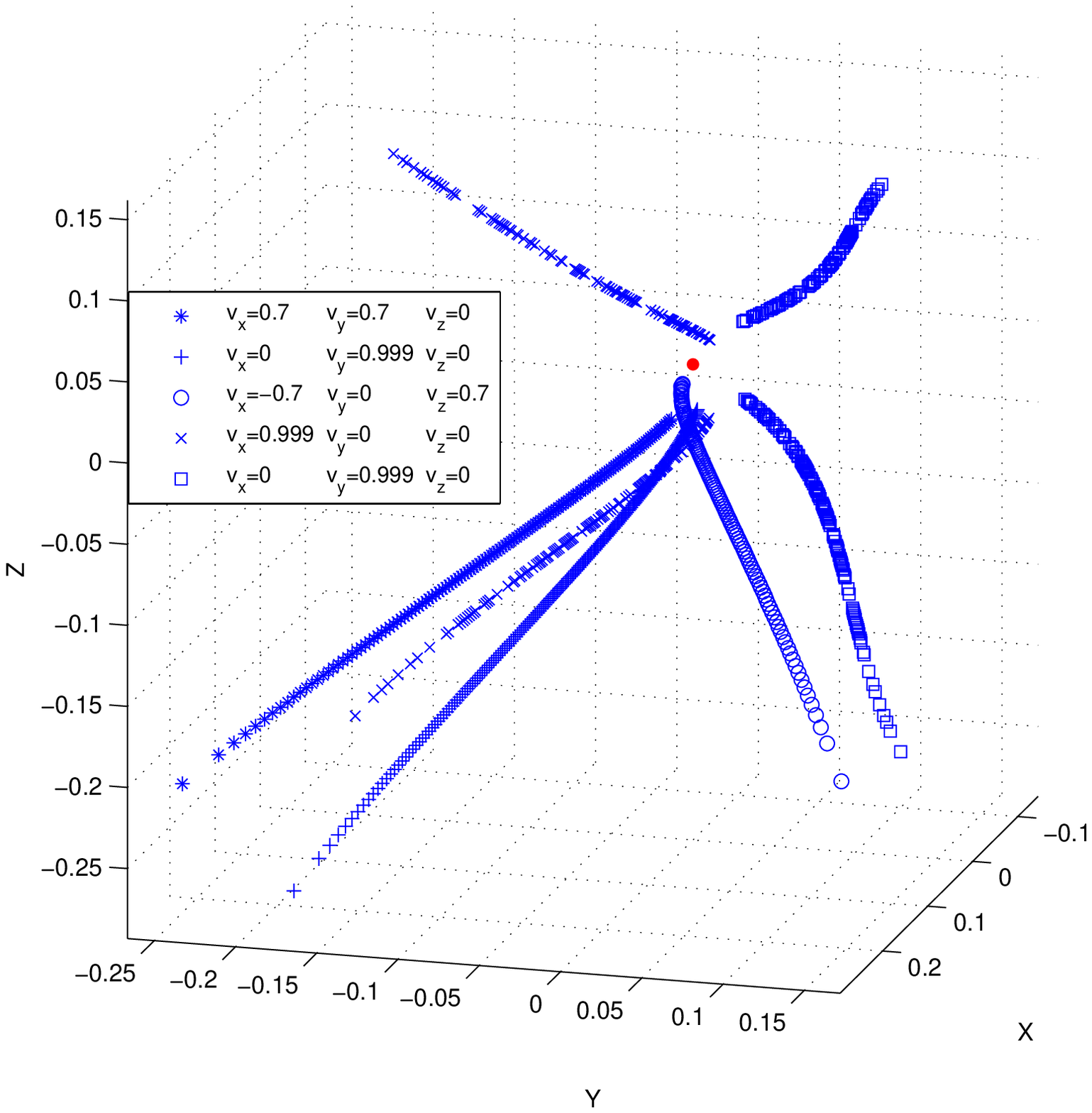}%{3Dnull_comb.eps}
\caption{The position of null points in a general case, plotted with respect to the mutual 
orientation of the black hole's drift velocity and the asymptotic magnetic field. The velocity 
components are 
indicated in the inset; from top to the bottom, the first three cases are for the oblique field 
($B_{x}>0$, $B_{x}=B_{z}$), while the remaining two examples show the aligned field 
($B_{z}>0$, $B_{x}=0$) for comparison. In the latter case the tracks of null points form 
pairs that are positioned symmetrically with respect to the equatorial ($z=0$) plane. 
The outer end of each track corresponds to the extreme spin, $a=1$. The spin decreases 
monotonically along each series of points towards the horizon. The rescaled radial coordinate 
has been employed, so the horizon corresponds to a single point in the origin of coordinates 
(depicted by the red dot).}
\label{fig8}
\end{figure}

A fully three-dimensional structure of magnetic lines 
develops outside the equatorial plane. In figure~\ref{fig4}
we observe a superposition of two essential effects. Firstly, 
the X-type structure of the magnetic null point persists also
outside the equatorial plane. Secondly,
however, the magnetic lines acquire a growing vertical 
component $B_z$ $(=B_\parallel)$, whereas the 
electric field passes through the magnetic null point and 
crosses the equatorial plane vertically. Such a structure
suggests that particles can be accelerated by the non-vanishing
electric field, and they can stream away from the neutral point.

It is interesting to investigate the location of magnetic null points as a function of
the black hole spin $a$ and velocity $v$ of the translatory motion. This question is
especially relevant in the context of the electromagnetic acceleration mechanisms of 
particles in the vicinity of rapidly rotating black holes (position of the magnetic null
point recedes from the horizon as the spin $a$ increases). We would like to identify the site of
magnetic reconnection and ask whether it is located outside the ergosphere or
inside it, depending on the black hole spin. The result is
shown in figure~\ref{fig5}, where we plot the radial distance of the magnetic null points 
as a function of spin for different values of the linear (boost) velocity $v_y>0$ ($v_x=v_z=0$).
Orientation of the velocity is set perpendicular to the asymptotic field, $B_{x}>0$,
$B_{z}=0$. This special configuration of a magnetic field (mutually orthogonal with respect 
to the drift velocity and the rotation axis) provides suitable conditions for the emergence 
of the null points. In the case of extreme spin we observe the occurrence 
of null points for velocity as low as $v_y\approx0.45$. On the other hand, 
rapid motion $v_y=0.999$ allows the null point formation also for a rather low spin 
about $a\approx0.2$. Both $a$ and $v$ parameters are essential to ensure the 
existence of null points, and they both help to drag the point farther out from the 
horizon. A combination of extreme spin $a\rightarrow M$ and fast 
motion $v_y=0.999$ brings the magnetic null to $r=2.203$ ($GM/c^2$). Increasing the 
velocity beyond this value does not move the neutral point any farther. On the 
other hand, for small values of spin and/or small velocity the null point does not 
show up at all.

Figure~\ref{fig6} completes information about the null point location in the
equatorial plane. Magnetic field is again specified to be asymptotically perpendicular 
to the rotation axis ($B_{x}>0$, $B_{z}=0$) and the black hole moves transversely 
with respect to both the magnetic field and the rotation axis ($v_y>0$, $v_x=v_z=0$).
The view of the equatorial plane in polar coordinates allows us to demonstrate how
the magnetic null points move gradually away from the horizon as the spin increases, while
they are dragged around the hole by its frame-dragging effect. For high velocity 
and large spin the null points emerge just outside the ergosphere boundary, 
whereas for lower values they are generally confined within the ergosphere.
We also observe that the distance of null points in oblique configurations 
of the system is generally smaller compared to the perpendicular case shown 
here, so we conclude that fig.\ \ref{fig6} shows the farthest location to which the 
magnetic null point can recede from the black hole horizon.

To conclude the discussion of the position of magnetic null points, 
in figure~\ref{fig7} we explore a general case when the spin of the black 
hole, the direction of its translatory motion, and the asymptotic magnetic field point all 
in different arbitrary directions. This case corresponds to the oblique system which lacks 
any special symmetry. Despite that, it is interesting to note that the magnetic null 
points again exist near the horizon (albeit out of the equatorial plane). Also in the 
oblique case, by raising the spin $a$ we locate the magnetic null 
points getting gradually farther away from the horizon (see figure~\ref{fig8}). 
The null points can be traced in analogy with those discussed previously
in the equatorial case (fig.\ \ref{fig6}), however, now the shape of the resulting 
tracks is more distorted and the null points positions are more difficult to find.
Because locating the null points of the oblique field does not seem to be possible
directly from the analytical form of the field components, we have searched the 
relevant region numerically.

Finally, it may be worth reminding the reader that the exact location of magnetic 
nulls depends on the choice of frame with respect to which we formulate the problem. We considered
a well-motivated physical frame and we checked that the null points occur also, e.g.,
in the frame of freely orbiting (Keplerian) observers, so the effect is relevant for the
behaviour of accreted matter.

\section{Discussion and Conclusions}
Rotation is an interesting attribute of cosmic black 
holes \citep{rn03}. In principle, black holes can be spun up close 
to extreme $a=1$ \citep{sb11}.
In stellar-mass black holes the spin is thought 
to be chiefly natal \citep{ms06}, whereas supermassive black holes 
in galactic cores can adjust their angular momentum by accretion, and the
outcome of evolution depends largely on the dominant mode of accretion, 
during their entire life-time \citep{fb11}. In both cases, the spin is an 
important characteristic, potentially allowing the efficient acceleration 
of matter. We showed that it can be also relevant directly for the onset 
of magnetic reconnection.

We discussed the structure of electromagnetic test-fields 
and the layered pattern of current sheets
that can be induced by the gravito-magnetic action. 
Neutral points of the magnetic field 
suggest that magnetic reconnection 
can occur. In the case of extreme rotation, $a=1$, the 
magnetic null points can occur just outside the ergosphere. 
We found that the null points exist also in a general
(oblique) case which lacks any special symmetry between the direction
of the magnetic field, velocity of the linear boost of the black hole, 
and the orientation of the spin axis. The proposed scenario can be 
astrophysically relevant in circumstances when the black
hole rotates and moves simultaneously across a magnetic flux tube. 
The origin of the magnetic flux tube is thought to be in currents 
flowing in the medium far from the 
black hole. We considered the limit of a magnetically dominated 
system in which the organised magnetic field dictates the 
motion to plasma; the opposite limit of a black hole moving through
a force-free plasma has been investigated by other authors
\citep[recently,][]{pg10}.

In this paper we considered an idealised situation, starting
from an electro-vacuum solution, assuming fast motion and rotation
of the black hole, and embedding it in an asymptotically uniform
magnetic field. Future simulations should clarify, whether
the astrophysically realistic effects of the moving 
black hole on the surrounding electromagnetic structure in its 
immediate neighborhood will be similar to those envisaged here. 
Despite the field-line structure in this paper being induced
purely by the action of frame-dragging, the exact choice of the 
projection tetrad is not very essential for the existence of 
magnetic layers. The choice of the physical frame
does affect the presence and the exact location of 
the magnetic neutral point, and the associated electric field
which accelerates charged particles away from the
neutral point. Although the exact location of the neutral point
varies with the model parameters, it always occurs very close to 
the black hole, where the frame-dragging is efficient. Therefore,
the point of particle acceleration has to be close
to horizon as well.

We note that all essential 
ingredients needed for the aforementioned mechanism to operate 
are met in standard circumstances that are believed to occur in
various cosmic objects. Namely, large scale magnetic fields
are frequently observed.
Furthermore, binary black holes are expected to exist in some 
nuclei of galaxies, where they are gravitational bound to each 
other and perform the orbital motion, while gradually inspiralling 
inwards; shortly before the merger, the motion speeds up 
significantly. Analytical models for this phase are not possible --
in general the spacetime structure has to be solved numerically, 
nevertheless, the magnetic structure of the organised 
field may arise quite generically in the close vicinity of such 
fast-moving black holes that are embedded within an external
magnetic field \citep{mp10}. Magnetic filaments exist in 
our galactic center and they suggest that the approximation of an 
asymptotically uniform organised magnetic field is worth 
investigating in this context.

\ack
We thank the Czech Science Foundation grant (No.\ 205/09/H033) and the Czech-US 
Cooperation Program (ME09036) for a continued support. We thank our referees
for constructive comments.

\end{document}